\documentclass[aps,prl,twocolumn,showpacs,superscriptaddress]{revtex4}%
\usepackage{amsfonts}
\usepackage{amsmath}
\usepackage{amssymb}
\usepackage{graphicx}
\usepackage{dcolumn}
\usepackage{bm}

\begin{document}
\title{Superfluid density in gapless superconductor CeCoIn$_5$}
\author{V. G. Kogan}
\email{kogan@ameslab.gov}
\affiliation{Ames Laboratory and Department of Physics \& Astronomy, Iowa State University,
Ames, Iowa 50011}
\author{R. Prozorov}
\email{prozorov@ameslab.gov}
\affiliation{Ames Laboratory and Department of Physics \& Astronomy, Iowa State University,
Ames, Iowa 50011}

\author{C. Petrovic}
\email{petrovic@bnl.gov}
\affiliation{Department of Physics, Brookhaven National Laboratory, Upton, NY 11973}



\begin{abstract}
 Temperature dependence of the London penetration depth $\lambda$ measured in single crystals of CeCoIn$_5$ is interpreted as caused by a strong pair-breaking scattering that makes the superconductivity in this compound  gapless. For a gapless d-wave superconductor, we derive $\lambda=\lambda_0/\sqrt{1-\left(T/T_c \right)^2}$ caused by combined effect of magnetic and non-magnetic scattering in excellent agreement with the data in the full temperature range and with the gapless s-wave case of Abrikosov and Gor'kov. We also obtain the slope of the upper critical field at $T_c$ that compares well with the measured slope.
\end{abstract}

\date{10 December 2007}
\maketitle

The heavy-fermion superconductor CeCoIn$_5$ is still under intensive scrutiny after its discovery in 2001  \cite{petrovic}.
This is  clean (the mean-free path exceeds by much the coherence length $\xi_{ab}\approx 80\,$\AA), nearly two-dimensional
(with small 3D pockets) \cite{Settai},  d-wave superconductor \cite{dwave}. In the normal phase for $T > T_c=2.3\,$K, this
material is a strong paramagnet \cite{Ames}. The main interest of the community has been focussed on  low temperatures and high
fields where the inhomogeneous Fulde-Ferrell-Larkin-Ovchinnikov (FFLO) phase is suspected to exist. In this work we are interested
in zero-field superfluid density and the $c$ directed upper critical field $H_{c2}$ near  $T_c$, the domain removed from complications of
FFLO and paramagnetic constrains. Understanding  the ground state properties is of  utmost importance for definite predictions
about existence of more complex phases such as FFLO.

Single crystals of CeCoIn$_5$ were grown from an In flux by combining stoichiometric amounts of Ce and Co with excess In in
an alumina crucible and encapsulating the crucible under vacuum in a quartz tube. Details of the crystal growth are described
elsewhere \cite{petrovic}. The crystals used were $1\times1\times0.2$ mm$^3$ and magnetic measurements showed practically no hysteresis.

The magnetic penetration depth was measured with a tunnel-diode resonator sensitive to changes in $\lambda$ of about 1
\AA. Details of the technique are described elsewhere \cite{Prozorov2000}. In short, properly biased tunnel diode
maintains self-resonating tank circuit on its resonant frequency $\omega \sim 14\,$MHz. A sample is inserted into the coil on a
sapphire rod. The changes in the effective inductance cause a shift in $\omega$  proportional to the real part of the dynamic magnetic
susceptibility. The system is calibrated by matching the $T$ dependent skin depth just above $T_c$. To probe $\lambda_{ab}$, a small ac
magnetic field ($\sim\,$20\,mOe) is applied along the c-axis, so that the screening currents are flowing in the $ab$ plane
\cite{Prozorov2000}.

\begin{figure}[tbh]
\begin{center}
\includegraphics[width=7.cm]{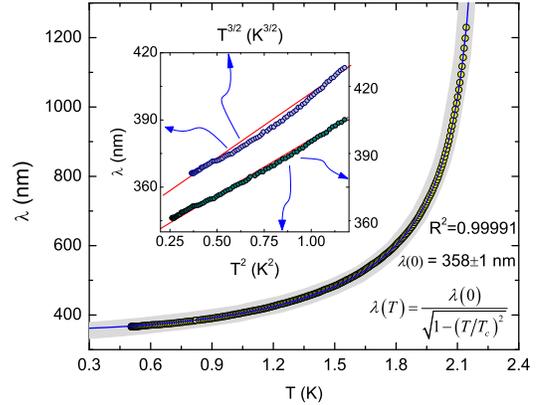}%
\caption{(Color online) The data on $\lambda(T)$ and the fit to Eq.\,(\ref{lam}) with  $\lambda(0)$ as the single  fitting parameter. Grey line:  data from Ref.\,\cite{Ozcan}. Inset: low temperature part of $\lambda(T)$ replotted vs. $T^{3/2}$ and $T^2$ for comparison.}
\label{fig1}%
\end{center}
\end{figure}

The London penetration depth $\lambda_{ab}(T)$ as a function of temperature shown in Fig.\,\ref{fig1}. With excellent
accuracy, the data from $0.5\,$K all the way to $2.1\,$K are described by
\begin{equation}
\lambda_{ab}= \lambda(0)\Big/\sqrt{1-t^2}  \,,\qquad
t= T/T_c \,.
\label{lam}
\end{equation}

In fact, our data are close to those reported in Ref.\,\cite{Ozcan} (the grey band  in Fig.\,\ref{fig1}); our interpretation, however,  is quite different.

We note  that  the superfluid density $ 1/\lambda^2$ deviates from the zero-$T$ value approximately as $T^2$  at low $T$'s, unlike both
the  s-wave exponentially weak behavior and  the clean d-wave linear $T$ dependence. On the other hand, we recall that Abrikosov
and Gor'kov (AG) in the seminal paper \cite{AG} on pair-breaking by magnetic impurities in isotropic s-wave materials found the
dependence (\ref{lam}) in the whole temperature range from 0 to $T_c$ for a strong spin-flip scattering when $T_c$ is suppressed to
nearly zero and the superconductivity is gapless.

This suggests that a similar situation takes place in CeCoIn$_5$, although  the  reason for the pair-breaking may not be
the spin-flip scattering on independent magnetic impurities of AG. Instead, it might be scattering on excitations of
the Kondo system of interacting local moments \cite{Kondo}. The situation  is far from being transparent as is shown by transport measurements with magnetic and
non-magnetic substitutions \cite{paglione}. Alternatively, the gaplessness might be caused by only a part of the Fermi surface being fully
gapped in multiband scenarios
\cite{Tanatar,BG}.

Since our technique only provides information on the change $\Delta \lambda (T)$, we compare our data with those of  Ref.\,\cite{Ozcan} where $\lambda (0)$ was estimated from the surface impedance data. By shifting our $\Delta \lambda (T)$ (circles in Fig.\,\ref{fig1}) we obtain one-to-one correspondence with the data of Ref.\,\cite{Ozcan} shown in Fig.\,\ref{fig1} by a grey band (the width is artificial - otherwise the data just collapse on top of each other). The solid line is the fit to Eq.\,(\ref{lam}) in the full temperature interval. The only fitting parameter, $\lambda(0)=358\,$nm, is larger than $ 281\,$nm obtained in Ref.\,\cite{Ozcan} from the low $T$ part of the data (shown in the inset). The fit is of a high precision reflected in the value of the ``coefficient of determination" $R^2=0.99991$ (for the perfect fit  $R^2=1$).

Inset to Fig.\,\ref{fig1} shows comparison of the data in the low-temperature domain with the form suggested in \cite{Ozcan}, $\Delta\lambda=\lambda-\lambda(0)\propto T^{3/2}$, and with the $\Delta\lambda\propto  T^{2}$, expected from our model (displaced vertically by 20 nm for clarity). Straight lines show clearly that the $T^{2}$ dependence describes the data better.

Long experience of dealing with pair-breaking effects reveals that all of them are described by the AG formalism, provided the dimensionless pair-breaking parameters are properly
defined for each particular case  \cite{Maki}. We therefore choose the AG scheme and characterize the scattering by two parameters
 \begin{equation}
  \rho  =
 \hbar/2\pi T_c\tau \,, \qquad \rho_m =  \hbar/2\pi
T_c\tau_m \,
 \label{rhos}
 \end{equation}
\noindent where $T_c$ is the critical temperature (not to confuse with hypothetical $T_{c0}$ of the material free of scattering), $1/\tau$ and $1/\tau_m$ are the transport and the  pair-breaking scattering rates. As mentioned, AG find $\lambda^{-2}\propto 1-t^2 $ for sufficiently strong pair-breaking in {\it dirty}  isotropic s-wave materials with  $\rho\gg \rho_m$. The material of interest here is a clean d-wave \cite{dwave};  we show below that this $T$ dependence holds in this situation as well.

Within microscopic theory, the penetration of weak magnetic fields into superconductors is evaluated by first solving for the unperturbed
zero-field  state and then treating effects of small fields as perturbations. Perhaps, the simplest for this task is the Eilenberger quasiclassical approach  \cite{E}.
Formally, it consists of equations for  Eilenberger functions $f({\bm  r},{\bm v},\omega),\,\,  f^{+} $, and $g$ which originate from Gor'kov's Green's functions integrated over the energy  near the Fermi
surface to exclude fast spatial oscillations on the scale $1/k_F$:
\begin{eqnarray}
{\bm v}{\bm\Pi} f&=&   \frac{2\Delta\, g}{ \hbar}-2\omega
f+   \frac{g}{\tau^-}\langle f\rangle -\frac{f}{\tau^+}\langle g\rangle \,,\label{Eil1}\\
-{\bm v}{\bm\Pi^*} f^+&=&   \frac{2\Delta^*\, g}{\hbar}-2\omega
f^+ +   \frac{g}{ \tau^-}\langle f^+\rangle -\frac{f^+}{ \tau^+}\langle g\rangle\,,\label{eil2}\\
1&=&g^2+ff^+ \,.\,\qquad \qquad \qquad \qquad \qquad \qquad
\label{eil3}
\label{normalization}
\end{eqnarray}
Here, ${\bm v}$ is the Fermi velocity, ${\bm  \Pi} =\nabla +2\pi i{\bm
A}/\phi_0$, $\phi_0$ is the flux quantum. $\Delta $ is  the order parameter  which may depend on the position ${\bm  k}_F$ at the Fermi surface (or on ${\bm v}$) in cases other than the isotropic
 s-wave; for a d-wave material with a simple cylindrical Fermi surface the order parameter can be written as
 \begin{equation}
\Delta=\Psi(\bm r,T)\Omega\,  ,\quad  \Omega =\sqrt{2}\cos\,2\varphi \,,
 \end{equation}
\noindent where $\varphi$ is the azimuthal angle on the Fermi cylinder and $\Omega$ is normalized to have $\langle \Omega^2\rangle=1$.

Further, $\omega$ are Matsubara frequencies defined by $\hbar\omega=\pi T(2n+1)$ with an integer $n$, $\langle ...\rangle$ denote  averages over the Fermi surface, and
\begin{eqnarray}
\frac{1}{\tau^\pm} =
\frac{1}{\tau }\pm\frac{1}{\tau_m} \,.
\end{eqnarray}

The system (\ref{Eil1})-(\ref{eil3}) should be complemented with the self-consistency equation for the order parameter and with an expression for the current density.   For the d-wave symmetry,  both magnetic and non-magnetic scattering suppress the critical temperature \cite{Openov}. The self-consistency equation in the form taking this into account, reads:
\begin{eqnarray}
  \ln \frac{T_{c}}{T}=\sum_{n=0}^\infty
\left (\frac{1}{n+1/2+\rho^+/2}-\frac{2\pi T}{\Psi}\Big\langle \Omega
f\Big\rangle\right )\,,
\label{self-cons}
\end{eqnarray}
where $\rho^+=\rho +\rho_m$. Finally, the current density expression completes the Eilenberger system:
\begin{eqnarray}
{\bm  j}=-4\pi |e|N(0)T\,\, {\rm Im}\sum_{\omega >0}\Big\langle {\bm v}g\Big\rangle\,;
\label{eil5}
\end{eqnarray}
 $N(0)$ is the total density  of states at the Fermi level per one spin.

Calculation of $\lambda(T;\tau,\tau_m)$ for arbitrary $\tau$'s is difficult analytically. However, for a strong $T_c$ suppression, the problem is manageable. We begin with uniform zero-field state for which $\rho+\rho_m$ is close to the
critical value where $T_c\to 0$;
  in this state   $f\ll 1$ and
$g\approx 1-f^2/2$ in the whole temperature range \cite{AG}. One can look for
 solutions of Eq.\,(\ref{Eil1}) as $f=f_1+f_2$ with $f_2\ll f_1$.
In the lowest approximation Eq.\,(\ref{Eil1}) yields:
\begin{eqnarray}
    2\Delta/\hbar-2\omega f_1+    \langle f_1\rangle/\tau^- -f_1 /\tau^+=0\,.
\label{1st}
\end{eqnarray}
Since $\langle \Delta\rangle=0$ for the d-wave materials, averaging of this equation over the Fermi surface gives $\langle f_1\rangle=0$ as well. Taking this into account we have
\begin{eqnarray}
    f_1= \Delta/\hbar\omega^+ \,,\qquad  \omega^+=\omega+1/
2 \tau^+  \,.
\label{f1}
\end{eqnarray}
The next approximation yields:
\begin{eqnarray}
    f_2=\frac{\Delta}{2\hbar^3\omega^{+\,3}}\left( \frac{\langle \Delta^2\rangle}{2\tau^+\omega^{+ }} -\Delta^2\right) \,.
\label{f1}
\end{eqnarray}

 Making use of $\Delta=\Omega\Psi$, $\langle \Omega^2 \rangle=1$,
$\langle \Omega^4 \rangle=3/4$ for a Fermi cylinder, we evaluate:
\begin{eqnarray}
\Big\langle \Omega (
f_1+f_2)\Big\rangle = \frac{\Psi}{\hbar\omega^+} -
\frac{\Psi^3(1-6\omega\tau^+)}{16\hbar^3\omega^{+\,4}\tau^+}\,.
\label{e100}
\end{eqnarray}
Substitute this in Eq.\,(\ref{self-cons}), express  the sums
in terms of  di-gamma functions, and utilize the asymptotic expansion $\psi(z+1/2)\approx \ln z+1/24 z^2$  for $z=\rho^+/2t\gg 1$:
\begin{eqnarray}
&&   -\ln t = \psi\left( \frac{\rho^+ }{2t}+\frac{1}{2} \right ) -
\psi\left( \frac{\rho^+ }{2}+\frac{1}{2} \right ) \nonumber\\
&&  -
\frac{3 \Psi^2}{32\pi^2T^2
}\left[\psi^{\prime\prime}\left( \frac{\rho^+}{2t}+\frac{1}{2}  \right
)-\frac{\rho^+ }{9t}
\psi^{\prime\prime\prime}\left( \frac{\rho^+}{2t}+\frac{1}{2} \right
)\right]\nonumber\\
&& \approx -\ln t+\frac{t^2-1}{6\rho^{+\,2}} +
\frac{ \Psi^2}{48\pi^2T_c^2\rho^{+\,2}}\,.
\label{e101}
\end{eqnarray}
Hence, we obtain:
\begin{equation}
\Psi^2=8\pi^2 \,(T_c^2-T^2)\,.
\label{Psi-d}
\end{equation}
This differs from the result  for isotropic s-wave
superconductors with magnetic impurities of nearly critical concentration,
   by a 4 times larger   pre-factor \cite{AG}. The ratio
\begin{equation}
 \Delta_{max}(0)/T_c =4\pi  \,
\label{ratio-d}
\end{equation}
is considerably larger than 1.764 of the BCS theory for isotropic
s-wave materials and 2.14 for the clean isotropic d-wave case. This ratio
estimated from the tunneling data on CeCoIn$_5$ is about 10
\cite{Laura Green}.

It is worth noting that even without magnetic scatterers, the
superconductivity in d-wave materials becomes gapless in the domain of
interest here with $\rho^+ \gg 1$. To see this, examine the
density of states $N(\epsilon)=N(0)\,{\rm Re}\,g(\hbar\omega\to
i\epsilon)$ using
$g=1-f^2/2=1-(\Delta/\hbar \omega^+)^2/2$:
\begin{eqnarray}
  \frac{N(\epsilon)}{N(0)}= 1-
\frac{2\Delta^2\tau^{+\,2}}{\hbar^2}\,\frac{1-\varepsilon^2}
{(1+\varepsilon^2)^2} \,,\quad \varepsilon=\frac{2\tau^+\epsilon}{\hbar}.
\label{N(e,gapless)}
\end{eqnarray}
Hence, at zero energy, $N(\epsilon)$ has a non-zero minimum (i.e., the superconductivity is gapless), whereas the maximum of $N(\epsilon)$ is reached at $\epsilon_{m}=   \hbar \sqrt{3}/2\tau^{+}$.


Weak supercurrents and fields leave the order parameter modulus unchanged,
but  cause the condensate to acquire an
overall phase $\theta({\bm  r})$. We therefore look for   perturbed solutions
in the form:
\begin{eqnarray}
\Delta  = \Delta _0 \, e^{i\theta},\,\,\,\,\,
f =(f_0  +f_1)\,e^{i\theta},\nonumber\\
f^{+} =(f_0  +f_1^+ )e^{-i\theta},\,\,\,\,\,
g =g_0 +g_1             ,
\label{perturbation}
\end{eqnarray}
where the subscript 1 denotes small corrections.  In the London
limit, the only coordinate dependence is that of the phase $\theta$, i.e.,
        $f_1 ,g_1 $ can be taken as ${\bm  r}$ independent.

The Eilenberger equations (\ref{Eil1})-(\ref{normalization}) provide the corrections among which we need only $g_1$:
\begin{equation}
g_1=\frac{i\hbar f_0^2 {\bm v}{\bm P}}{2({\tilde \Delta} f_0+
     \hbar{\tilde\omega}g_0)}\approx \frac{i  f_0^2 {\bm v}
{\bm P}}{2    \omega^+  }\,.
\label{g1}
\end{equation}
Here ${\bm  P}= \nabla\theta+ 2\pi{\bm  A}/\phi_0\equiv 2\pi\,
{\bm  a}/\phi_0$ with the ``gauge invariant vector potential"  ${\bm  a}$,
and
        \begin{eqnarray}
{\tilde \Delta} = \Delta + \hbar
\langle f\rangle/2\tau^- \,, \quad
     {\tilde\omega} = \omega  +  \langle g\rangle/2\tau^+
 \label{w-tilde1}\,.
\end{eqnarray}
In the case of interest, $f_0\approx\Delta/\hbar\omega^+\ll 1$, and the
denominator in Eq.\,(\ref{g1}) is taken in the lowest order. We
now substitute $g_0+g_1$ in the   current  (\ref{eil5})  and
compare the result with   $4\pi
j_i/c=-(\lambda^2)_{ik}^{-1}a_k$ to obtain:
\begin{equation}
 \lambda_{aa}^{-2}= \frac{32\pi e^2N(0)
  v^2 }{ c^2\rho^{+\,2}}\,(1-t^2).  \label{lambda-d-tau}
\end{equation}

Using the data of Fig.\,\ref{fig1} with  estimates for $N(0)$ and $v$ taken from
\cite{Ames}, we obtain $\rho^+\approx 8$.
Estimates of \cite{dwave} yield $\rho^+\approx 5$.
Hence, the statement that CeCoIn$_5$ is gapless is
not only in excellent agreement with  the $T$ dependence of $\lambda$, but the scattering parameter $\rho^+$ is sufficiently large  which is needed for our model to hold.


 There are quite a few reports of $H_{c2}(T )$ for CeCoIn$_5$, see, e.g., \cite{Hc2 data}. Our data are shown in Fig.2. For a strong
pair-breaking model with  fixed scattering parameters $H_{c2}\propto
(1-t^2)$ \cite{AG} that is clearly different from the experimental behavior. In a strong paramagnet such as CeCoIn$_5$, the magnetic
scattering rate (and  $\rho_m$)  may itself depend on the applied
field, making our model with a constant $\rho_m$ inapplicable {\it per se} along the whole $H_{c2}(T)$ curve. However, near $T_c$ where  $H_{c2}\to 0$, we expect the model to hold. We, therefore, evaluate only $H_{c2}(T)$ near
$T_c$; in other words, we  derive the  Ginzburg-Landau equation  containing
the coherence length $\xi$ for the gapless case with a strong pair-breaking.

Near $T_c$, we  look for a solution $f$ of  Eq.\,(\ref{Eil1})
as an expansion in two small parameters
$$\frac{\Delta}{\hbar\omega^+}\sim\frac{\Delta}{T_c}
\sim  \delta t ^{1/2},\,\, {\rm and}\,\,\, \frac{{\bm
v}{\bm\Pi}\Delta}{\hbar\omega^{+\,2}}\sim \frac{
\Delta}{T_c\xi}\sim  \delta t $$
where $\delta t=1-T/T_c$; smallness of the second parameter comes from the
``slow variation" requirement. We obtain after simple algebra:
\begin{eqnarray}
    f =\frac{\Delta}{ \hbar \omega^{+ }}- \frac{{\bm
v}{\bm\Pi}\Delta}{2\hbar\omega^{+\,2}} +{\cal O}( \delta t ^{3/2})\,.
\label{f1}
\end{eqnarray}
Further, the self-consistency Eq.\,(\ref{self-cons}) which now reads
as
\begin{eqnarray}
\frac{\Psi}{2\pi T_c}\,\delta t=\sum_{\omega>0}
\left (\frac{\Psi}{\hbar\omega+\pi T\rho^+}- \Big\langle
\Omega f\Big\rangle\right ),
\label{eil4dd}
\end{eqnarray}
should be taken into account.  To evaluate  $ \langle
\Omega f \rangle$, we substitute $g=1-ff^+/2$ in
Eq.\,(\ref{Eil1}), multiply it by $\Omega/\omega^+$, and take average over
the Fermi surface:
\begin{eqnarray}
\frac{1}{2\omega^+}\Big\langle{\Omega\bm v}{\bm\Pi} f\Big\rangle &=&
 \frac{\Psi }{\hbar \omega^{+ }}-
\Big\langle\Omega f\Big\rangle  \nonumber\\
&-&  \frac{3\Psi |\Psi |^2}{8\hbar^3\omega^{+\,3}}
  +  \frac{\Psi |\Psi
|^2}{4\hbar^3\omega^{+\,4}\tau^+}\, .
\label{eil1ddd}
\end{eqnarray}
Writing the last two terms, we can take $f$ in the lowest order;  also we
make  use of $\Delta=\Psi\Omega$, $\langle
\Omega^2 \rangle=1$, and of   $\langle
\Omega^4 \rangle=3/4$.
On the left-hand side we have:
\begin{eqnarray}
\frac{1}{2\hbar\omega^{+\,2}}\Big\langle{\Omega^2\bm v}{\bm\Pi}
\Psi\Big\rangle + \frac{1}{4\hbar\omega^{+\,3}}\Big\langle \Omega^2({\bm
v}{\bm\Pi})^2
\Psi\Big\rangle \nonumber\\
=\frac{1}{4\hbar\omega^{+\,3}}\Big\langle \Omega^2v_iv_k\Big\rangle
 \Pi_i\Pi_k
\Psi\,;
\label{lhs}
\end{eqnarray}
summation over the repeated indices is implied.

We now sum up Eq.\,(\ref{eil1ddd}) over $\omega$ and take $\sum\langle
\Omega f \rangle$ from the self-consistency Eq.\,(\ref{eil4dd}).
All sums obtained are expressed in terms of di-gamma functions, for which the
large argument asymptotics can be used.  We are interested in
the field along the $c$ crystal axis, whereas the
plane $ab$ can be taken as isotropic. After a straightforward algebra one
obtains the GL equation in the standard form
\begin{equation}
-\xi^2\Pi^2\Psi=\Psi(1-|\Psi|^2/\Psi_0^2)
\end{equation}
with
\begin{eqnarray}
 \xi^2=\frac{3\hbar^2v^2}{32 \pi^2T_c^2\delta t}\,,\quad  \Psi_0^2=
16\pi^2T_c^2\delta t\,.
\label{xi}
\end{eqnarray}
We note  that   $\Psi_0^2$ is the zero-field
 order parameter of Eq.\,(\ref{Psi-d}) obtained there by a different method.

Although the scattering parameters do not enter explicitly   the coherence length, they  affect $\xi$
through  $T_c(\rho^+)$.
Equation (\ref{xi}) shows that the scattering, whatever it is, {\it enhances }$\xi$. In other words, in the gapless d-wave regime, the effect of scattering upon the coherence length is opposite  to the familiar s-wave situation where $\xi$ is {\it suppressed} by scattering. When the scattering approaches the critical value for which $T_c\to 0$, $\xi$ diverges.
\begin{figure}[htb]
\begin{center}
\includegraphics[width=7.cm]{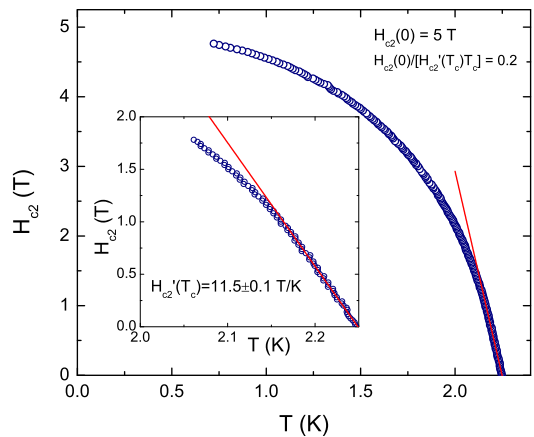}%
\caption{(Color online) The upper critical field $H_{c2}(T)\parallel \bm c$. Inset shows the domain near $T_c$ with an estimate of the slope.}%
\label{fig2}%
\end{center}
\end{figure}
The slope of the upper critical field at $T\to T_c$  follows:
\begin{eqnarray}
 H_{c2}^\prime(T_c)=-\frac{16\pi \phi_0T_c k_B^2}{3\hbar^2v^2}\,,
\label{slope}
\end{eqnarray}
where the temperature is given in Kelvins.  Hence, the slope decreases when scattering is intensified. This unusual behavior is confirmed by the data on $H_{c2}$ for low concentration La doping of Ce$_{1-x}$La$_{x}$CoIn$_5$    \cite{Ames}. Moreover, we  note that the Fermi velocity
  can now be expressed in terms of the measured slope and $T_c$. This yields $v=1.1\times 10^6\,$cm/s, the value between $v=2\times 10^6\,$cm/s obtained using the heat capacity and resistivity data in Ref.\,\cite{Ames} and $v=0.77\times 10^6\,$cm/s as estimated in Ref.\,\cite{dwave}.


 To conclude, we show that the data on the $T$
dependence of the penetration depth in the full  temperature range and on the upper critical field
near $T_c$ strongly support the notion that the superconductivity in CeCoIn$_5$ is gapless.

We thank Paul Canfield and Sergey Bud'ko for many useful
discussions. The work at the Ames Laboratory is supported by the
Office of Basic Energy Sciences of the US Department of Energy under
the Contract No. DE-AC02-07CH11358. R.P. acknowledges  support
of Alfred P. Sloan Foundation.

\end{document}